\documentclass[useAMS,usenatbib]{mn2e}
\usepackage{psfig}
\usepackage{graphicx}

\title[Simple Stellar Population Models as probed by ESO~121--SC03]
{Simple Stellar Population Models as probed by the Large Magellanic Cloud
Star Cluster ESO 121--SC03}

\author[Y. Xin et al.]{Y. Xin$^{1, 2, 3}$\thanks{E-mail:
y.xin@sheffield.ac.uk}, L. Deng$^{2}$, R. de Grijs$^{1, 2}$,
A. D. Mackey$^{4}$, and Z. Han$^{3}$\\
$^{1}$Department of Physics and Astronomy, The University of
Sheffield, Hicks Building, Hounsfield Road, Sheffield S3 7RH\\
$^{2}$National Astronomical Observatories, Chinese Academy of
Sciences, Beijing 100012, P. R. China\\
$^{3}$National Astronomical Observatories/Yunnan Observatory,
Chinese Academy of Sciences, Kunming, Yunnan 650011,
P. R. China\\
$^{4}$Institute for Astronomy, University of Edinburgh, Royal
Observatory, Blackford Hill, Edinburgh, EH9 3HJ}

\begin{document}

\date{Accepted ---. Received ---; in original form ---}

\pagerange{\pageref{firstpage}--\pageref{lastpage}} \pubyear{2007}

\maketitle

\label{firstpage}

\begin{abstract}
The presence of blue straggler stars (BSs) in star clusters has
proven a challenge to conventional simple stellar population (SSP)
models. Conventional SSP models are based on the evolution theory of
single stars. Meanwhile, the typical locations of BSs in the
colour-magnitude diagram of a cluster are brighter and bluer than
the main sequence turn-off point. Such loci cannot be predicted by
single-star evolution theory. However, stars with such properties
contribute significantly to the integrated light of the cluster. In
this paper, we reconstruct the integrated properties of the Large
Magellanic Cloud cluster ESO~121--SC03, the only cluster populating
the well-known age gap in the cluster age distribution, based on a
detailed exploration of the individual cluster stars, and with
particular emphasis on the cluster's BSs. We find that the
integrated light properties of ESO~121--SC03 are dramatically
modified by its BS component. The integrated spectral energy
distribution (ISED) flux level is significantly enhanced toward
shorter wavelengths, and all broad-band colours become bluer. When
fitting the fully integrated ISED of this cluster based on
conventional SSP models, the best-fitting values of age and
metallicity are significantly underestimated compared to the true
cluster parameters. The age underestimate is $\sim40$ per cent if we
only include the BSs within the cluster's half-light radius and
$\sim60$ per cent if all BSs are included. The corresponding
underestimates of the cluster's metallicity are $\sim30$ and
$\sim60$ per cent, respectively. The populous star clusters in the
Magellanic Clouds are ideal objects to explore the potential
importance of BSs for the integrated light properties of more
distant unresolved star clusters in a statistically robust manner,
since they cover a large range in age and metallicity.

\end{abstract}


\section{Introduction}
Even using improved observational technologies, the number of
galaxies with resolved stellar contents is limited. Comparison of
the observed integrated spectral properties with models of
``simple'' stellar populations (SSPs; single-age, single-metallicity
stellar populations), i.e. using the so-called population synthesis
technique, is therefore a practical approach to studying the
formation and evolution processes in unresolved galaxies and their
components. As the basic building blocks of the evolutionary
population synthesis (EPS) method, conventional SSP models are
generally constructed based on the evolution theory of single stars.
(We specify the conventional SSP models throughout this paper as the
theoretical SSP models based on single-star evolution theory.) Thus
far, accuracy assessments of the existing suites of conventional SSP
models have not yet adequately considered a potentially significant
problem revealed by observations of star clusters.

By assuming that all the original member stars within a cluster are
born at the same time from the same progenitor molecular cloud, so
that they have the same age and metallicity, star clusters are
generally considered as the closest counterparts in the real world
to idealised SSPs. The major differences between a star cluster and
a conventional model SSP of the same age and metallicity can be
illustrated by fitting the observed colour-magnitude diagram (CMD)
of the cluster with a theoretical single-star isochrone. Most of the
member stars can be well fitted by the isochrone in terms of their
positions in the CMD. The number distributions along the isochrone
can be retrieved via the adopted stellar initial mass function
(IMF), and the number of low-mass member stars can be conserved in
the presence of evaporation due to dynamical evolution. Then, the
most prominent difference should be attributed to those member stars
``straggling'' away from the isochrone. Yellow and red stragglers
(Protegies~Zwart et al. 1997a,b) are not considered in the work,
because they are comparatively rare and less luminous than blue
stragglers (BSs). The underlopers and faint stars bluer than the
main sequence (MS) are not bright enough to be important. Therefore,
special attention has only been focussed on the BSs in our previous
work (Deng et al. 1999; Xin \& Deng 2005, hereafter XD05; Xin, Deng
\& Han 2007, hereafter XDH07), in view of their potential
non-negligible influence on the integrated properties of star
clusters.

BSs have been widely observed in stellar systems of all scales and
complexities, as shown by, e.g. Ahumada \& Lapasset (2007) for
Galactic open clusters (OCs), Piotto et al. (2004) for Galactic
globular clusters (GCs), Carney et al. (2005) for the Galactic field
stellar population, Johnson et al. (1999) for Large Magellanic Cloud
(LMC) clusters, Alcaino et al. (2003) for Small Magellanic Cloud
(SMC) clusters, and Mapelli et al. (2007) for dwarf spheroidal
galaxies. Considering their general locations in the CMD, i.e.,
brighter and bluer than the MS turn-off point, BSs are remarkably
hotter than the most luminous ``normal'' MS stars. Typically for old
Population I star clusters, when the red clump giants (RCGs) are
populated instead of the blue horizontal branch (HB), BSs can
significantly enhance the cluster' integrated spectrum at
ultraviolet (UV) and blue wavelengths -- the modifications may lead to
the star cluster being predicted as younger or of lower metallicity
based on conventional SSP models, and therefore cause uncertainties
when simply applying the conventional SSP models in EPS studies
(XD05; XDH07). Meanwhile, the most likely formation mechanisms of
BSs are all related to stellar interactions, including mass transfer
in close binaries (e.g., Tian et al. 2006; and references therein)
and stellar collisions in high-density regions (e.g., Sills et al.
2005; and references therein). This provides the theoretical support for
this empirical work, since the consequences of the products of
stellar interactions are not included in the conventional SSP
models.

In order to empirically correct the conventional SSP models for the
presence of luminous BSs, we construct the integrated spectrum of an
SSP (i.e., a realistic star cluster) by analysing the individual
stars, after careful consideration of their membership probability.
We assume that the stellar population of a cluster is composed of
two components: (i) all member stars that can be well fitted by a
single-star isochrone represent a conventional SSP for the cluster's
real age and metallicity (e.g., SSP models from Bruzual \& Charlot
2003; hereafter BC03); and (ii) BSs are responsible for the
modifications to the conventional SSPs, and their contributions are
considered by including the individual spectra of BSs. By combining
these two components we may obtain a better approximation to the
true SSP corresponding to the cluster of interest (see for details
XD05 and XDH07).

In our previous work (XD05; XDH07), a sample of 100 Galactic OCs
spanning an age range $\in[0.1, 10]$ Gyr and a range in metallicity,
$Z\in[0.005, 0.035]$ (where solar metallicity is Z$_\odot = 0.020$),
was used to detect BS contribution semi-empirically. The general
results from our Galactic OC studies show that the integrated
spectral properties of the sample clusters are dramatically modified
by their BS components. A preliminary assessment of the
uncertainties inherent to the conventional SSP models was made.
Using either spectra or broad-band colours, the resulting ages
and/or metallicities will be underestimated significantly:
conservatively, the age underestimates are $\sim50$ per cent, as are
the likely underestimates of the metallicity (although they are less
accurately constrained because of our lower resolution in
metallicity).

The results from the Galactic OCs clearly showed the non-negligible
contributions of BSs to the integrated properties of SSPs. However,
the numbers and distributions of BSs in OCs are affected by
significant stochastic uncertainties (Ahumada \& Lapasset 1995,
2007). To make the results and conclusions more reliable and
convincing, we expand our working sample by including star clusters
in the Magellanic Clouds (MCs). MC clusters hold significant
advantages over Galactic OCs: (i) they span a wider range in both
age and metallicity, and (ii) they are significantly more massive
than their counterparts in the Galaxy, and therefore provide better
statistics on an individual cluster basis (e.g., Mackey \& Gilmore
2003a,b; de~Grijs \& Anders 2006).

In this paper, the LMC cluster ESO~121--SC03 is analyzed using its
integrated spectral properties. This serves as the first example to
detect BS contributions to the conventional SSP models using a
massive intermediate-age cluster in a low-metallicity environment,
based on a novel approach and better statistics than we were
afforded by our use of Galactic OCs. ESO~121--SC03
($\alpha_{2000}=06^{\rm h}02^{\rm m}01^{\rm s}.36$,
$\delta_{2000}=-60^\circ31'22''.6$) is a distant northern LMC
cluster, lying at a projected angular separation of $\sim10^\circ$
from the LMC centre. It is described as a ``unique'' LMC cluster by
Mackey, Payne \& Gilmore (2006; hereafter MPG06), because it is the
only known cluster to lie in the LMC age gap. A significant number
of previous studies (e.g., Mateo et al. 1986; Bica et al. 1998;
MPG06) claim an absolute age of ESO~121--SC03 in the range of $8-10$
Gyr. Mateo et al. (1986) obtained [Fe/H]~$=-0.9\pm0.2$ combined with
a reddening of $E(B-V)=0.03$ mag. MPG06 derive
[Fe/H]~$=-0.97\pm0.01$ and $E(V-I)=0.04\pm0.02$ for the cluster.
MPG06 also mark a region in the CMD used to define BS candidates in
the cluster, but they do not study the BS population in detail.

We present our photometric data reduction steps in Section 2. In the
subsequent sections we present the details of the model construction
and our main results. The CMD and the best-fitting isochrone of the
cluster are discussed in Section 3. The identification of BSs in the
CMD is described in Section 4. A discussion of the contributions of
BSs to the integrated spectral properties of the cluster is presented
in Section 5. In Section 6, the synthetic integrated spectral energy
distribution (ISED) of the cluster is fitted with conventional SSP
models in order to assess the uncertainties. Finally, a summary and
the conclusions of this study are presented in Section 7.

\section{Data reduction}

We take advantage of the accurate photometric data taken with the
Advanced Camera for Surveys (ACS) Wide Field Channel (WFC) on board
the {\sl Hubble Space Telescope (HST)} as part of the {\sl HST} Cycle
12 snapshot survey of MC star clusters (programme 9891, PI
G. Gilmore). The ACS WFC consists of two 2048$\times$4096 pixel CCDs
with a scale of $\sim 0.05$ arcsec pixel$^{-1}$, separated by a gap of
$\sim 50$ pixels, and approximately covers a field of view of 202
$\times$ 202 arcsec$^2$. The frame was taken in each of two filters,
F555W and F814W. Exposure times were 300s and 200s, respectively.
Further instrumental and observational details can be found in series
of publications based on this programme, e.g., Mackey \& Gilmore
(2004), MPG06, and Mackey \& Broby~Nielsen (2007).

We retrieved the FITS files from the STScI data archive. The
observations were reduced using the STScI reduction pipeline, i.e.,
they have had bias and dark current frames subtracted and are divided
by flat-field images. The photometry was performed with the DOLPHOT
software (Dolphin 2000), specifically the ACS module. Before we
performed the photometry, we first prepared the images using the
DOLPHOT tasks \textit{acsmask} and \textit{splitgroups}. These two
packages mask out all bad pixels in the images and then split the
multi-image STScI FITS files into a single FITS file per chip,
respectively. We then used the main DOLPHOT routine to make
photometric measurements on the pre-processed images, using the F814W
drizzled frame as the position reference. All running parameters were
set to the recommended values in the DOLPHOT manual. The output
photometry was obtained in the VEGAMAG system and corrected for
charge-transfer efficiency degradation. Photometric calibrations and
transformations were done following Sirianni et al. (2005).

To obtain a high-quality CMD of the cluster, we used three parameters
from DOLPHOT to filter the photometric results. The ``sharpness''
measures how broad the intensity profile of a detected object is
relative to the point-spread function (PSF). It is zero for a
perfectly fit star, positive for an object that is too sharp (i.e.,
perhaps a cosmic ray), and negative for an object that is too broad
(i.e., perhaps a blend, cluster, or background galaxy). The
``crowding'' parameter is given in magnitudes, and measures how much
brighter the object would have been measured had nearby objects not
been fit simultaneously. For an isolated star the ``crowding'' value
is zero. The $\chi^2$ parameter represents the quality of the PSF
fitting. In this paper, we selected only objects with $-0.3 \leq$
sharpness $\leq 0.3$, crowding $\leq 0.5$ mag, and $\chi^2 \leq 0.25$
in both frames. Meanwhile, we only kept objects classified as good
star (object type 1) and star errors of types 1--7 by DOLPHOT, which
are referred to as ``usable'' in the DOLPHOT manual.

\begin{figure}
\includegraphics[width=8cm]{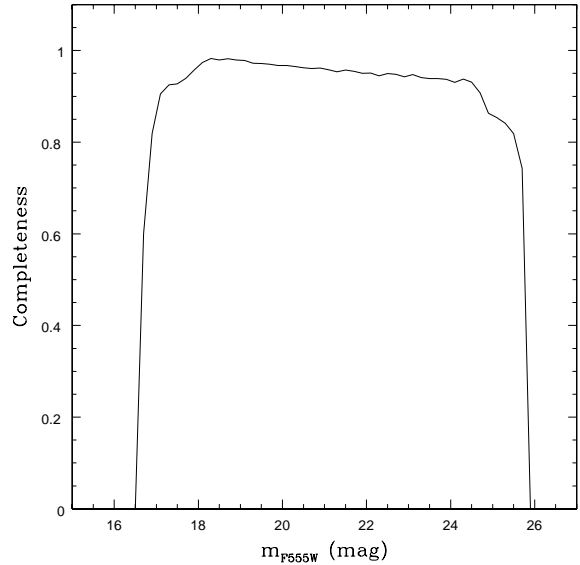}
\caption{Completeness function of ESO~121--SC03. The function has been
integrated over position and colour as a function of $m_{\rm F555W}$.}
\label{fig1}
\end{figure}

After following through this procedure, 2824 objects were detected in
the field of ESO~121-SC03. In order to evaluate the completeness of
the data, DOLPHOT was employed again, in artificial-star mode. We
generated $\sim 10^6$ fake stars, which were created with the DOLPHOT
task \textit{acsfakelist}. The limits of the artificial stars were set
to 16.0--28.0 mag in magnitudes and $-0.50-2.00$ mag in colours in the
CMD. The fake stars were binned in four-dimensional space: 200 pixels
in {\it x}- and {\it y}- positions, respectively, 0.2 mag in
brightness and 0.25 mag in colour. The completeness function of the
cluster is shown in Fig.~\ref{fig1}, integrated over position and
colour, as a function of $m_{\rm F555W}$. Completeness declines for
$m_{\rm F555W} \leq 18$ mag because we do not have any stars of that
brightness in the CMD.

\section{Colour-magnitude diagram}

\begin{figure}
\includegraphics[width=8cm]{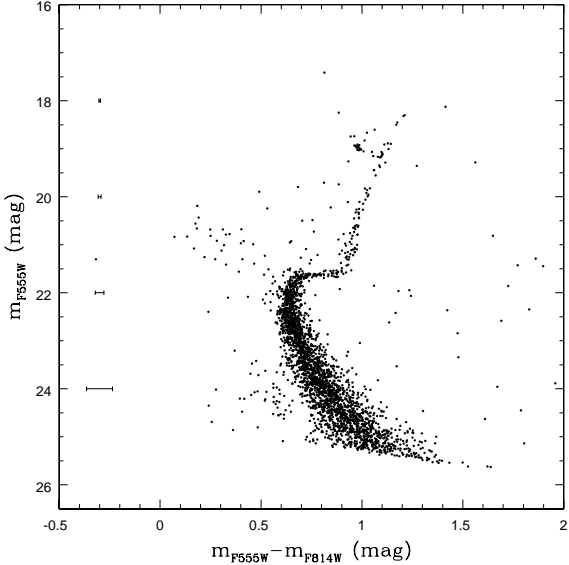}
\caption{Cleaned CMD of ESO~121--SC03. The CMD contains 2824
detections. The photometric data is plotted in the VEGAMAG
system.}\label{fig2}
\end{figure}

The final, cleaned CMD of ESO~121--SC03 in $m_{\rm F555W}$ and
$m_{\rm F814W}$ magnitudes is presented in Fig.~\ref{fig2}. It is
clearly shown that the CMD is well defined and contains little field
contamination. The perfectly represented evolutionary stages, e.g.,
MS, red-giant branch (RGB), and RCG, indicate the high accuracy of
the photometry and the validity of the reduction procedure. The CMD
reaches more than 3 magnitudes below the Main sequence turn-off
(MSTO). All these aspects help us to perform a detailed and highly
accurate statistical study of the cluster.

As described in Section 1, the synthetic ISED of a cluster is
constructed by assuming that the entire stellar population of the
cluster can be decomposed into two components: one accounts for all
member stars that are well fitted by a given isochrone (of single
stars) for the appropriate cluster age and metallicity; the other
includes only the cluster's BS population, which is therefore referred
to as the BS component. All other member stars deviating from the
fitted isochrone can be neglected because of their low luminosity and
relatively small number. More details justifying this approach can be
found in XD05.

\begin{figure}
\includegraphics[width=8cm]{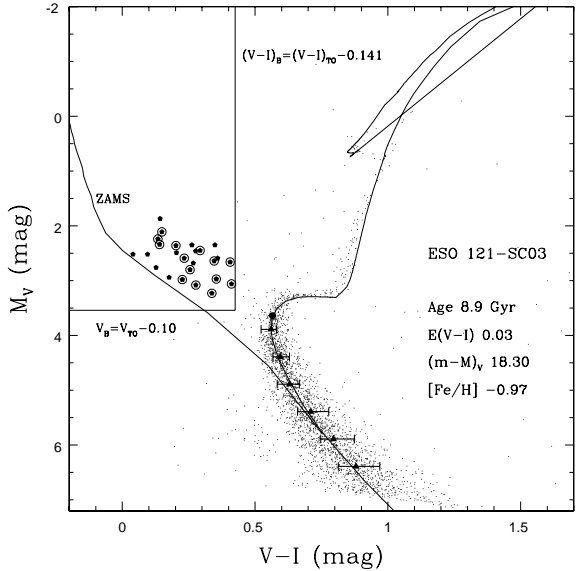}
\caption{Cleaned CMD for ESO~121--SC03. The photometry has been
transformed from the VEGAMAG system to Johnson--Cousins $V$ and $I$
magnitudes. The best-fitting Padova2000 isochrone is overplotted.
The corresponding fit parameters are included in the figure legend.
In the CMD, we define a region that we will use for the
identification of the cluster's BS population. The ZAMS is the
Padova1994 isochrone at log(age yr$^{-1}$) = 6.60 and
[Fe/H]~$=-0.97$ dex. The solid bullet is the cluster's MSTO. The
solid triangles represent the MSRL of the cluster. Pentagons are
blue stragglers. See the text (Sections 3 and 4) for
details.}\label{fig3}
\end{figure}

To follow the same procedure for ESO~121--SC03, we first need to
find the isochrone that best matches the observed features of the
cluster's CMD, and use this to derive accurate physical parameters,
including the cluster's age, metallicity, colour excess and distance
modulus. These parameters are very important for our construction of
the ISED of the SSP component and for defining the BS population.
The photometry in the VEGAMAG system is transformed into
Johnson--Cousins \textit{V} and \textit{I} magnitudes following the
prescription of Sirianni et al. (2005), and subsequently the new CMD
in $V,(V-I)$ is used to find the best-fitting isochrone. The fit
quality and corresponding details are displayed in Fig.~\ref{fig3}.

As the first step to identify the best-fitting isochrone, we attempted
to identify the MSTO in the CMD by counting the number of stars in
bins of magnitude (0.01 mag) and colour (0.001 mag) in the turn-off
region; the intersection of two bins with maximum stellar numbers, one
in magnitude and the other in colour, is regarded as the cluster's
MSTO. The resulting MSTO locus is at $(V-I)_{\rm TO} = 0.606\pm0.005$
mag and $V_{\rm TO} = 21.94\pm0.01$ mag. Next, the same procedure is
adopted to identify the main sequence ridge line (MSRL) in the CMD:
the numbers of stars are measured in magnitude bins of 0.5 mag, from
the MSTO to 3 magnitudes below the MSTO. In Fig.~\ref{fig3}, the
calculated MSTO is marked as the solid bullet, and the MSRL is marked
by solid triangles.

We used the Padova2000 theoretical isochrones (Girardi et al. 2000) to
fit the observed CMD. The original Padova2000 isochrones were
interpolated to [Fe/H]~$=-0.97$ dex (the value given by MPG06). In
total nine points, including the six MSRL points, the bottom of the
RGB, the mean location of the RCG, and the point on the RGB at the
level of the intermediate magnitude between the points of the RGB
bottom and the mean RCG locus, were used as fiducial points to
evaluate the fit quality. Next, we calculated the standard deviation
($\sigma$) for the offsets between the nine fiducial points in the CMD
and the isochrone, in order to obtain the minimum $\sigma$ value. The
isochrone fitting also offered us estimates for the apparent distance
modulus, $(m-M)_V$, and the foreground extinction suffered by the
cluster.

Given the above results and discussion, we adopted the metallicity of
[Fe/H]~$=-0.97$ dex, and we found a best-fitting isochrone of
age~=~8.9$^{+1.1}_{-0.3}$ Gyr, with $(m-M)_V=18.30\pm0.06$ mag and
$E(V-I)=0.03\pm0.01$ mag. These values will be used in the remainder
of this paper.

Our measurements for ESO~121--SC03 are in good agreement with previous
results. We derived that the age of the cluster is $8.9^{+1.1}_{-0.3}$
Gyr, which is entirely consistent with the age estimates of MPG06, who
provided an age range of $8.3-9.8$ Gyr for ESO~121--SC03. Similarly,
Mateo et al. (1986) found that this cluster is $10\pm2$ Gyr or $8\pm2$
Gyr old if the LMC distance modulus is 18.2 or 18.7 mag,
respectively. Bica et al. (1998) gave an age of 8.5 Gyr for the
cluster from Washington photometry.  Besides the similar age value,
our result confirm the conclusion of MPG06 and Mateo et al. (1986)
that ESO~121--SC03 is the only known cluster lying within the LMC age
gap. The LMC has a unique cluster formation history where nearly all
of its star clusters were formed either $\sim13~Gyr$ ago or less than
$\sim3~Gyr$ ago (Da~Costa 1991; Geisler et al. 1997 and references
therein; see also Parmentier \& de Grijs 2007). As suggested by
numerical simulations (e.g., Bekki et al. 2004), the origin of the age
gap is associated with the dynamical interaction between the LMC and
the SMC about $3-4$ Gyr ago. However, no theoretical studies have
explained the existence of ESO~121--SC03. Bica et al. (1998)
considered that this cluster may have originated from a recent
accretion of dwarf galaxies, but Dirsch et al. (2000) concluded that
the accretion of ESO~121--SC03 is not necessary, since they found that
this cluster has a similar age-metallicity relation as the LMC field
stars. In this respect, ESO~121--SC03 may reveal important information
about the formation and evolution history of the LMC.

The $(m-M)_V$ and $E(V-I)$ values for ESO~121--SC03 were determined
based on the offsets in, respectively, magnitude and colour required
to align the MSTO of the isochrone with that of the observed
CMD. Thus, it is worth considering the consistency between our results
and those from other work based on different methods. There are few
estimates of the colour excess for ESO~121--SC03. Both Mateo et
al. (1986) and Bica et al. (1998) adopted $E(B-V)=0.03$ mag from the
Burstein \& Heiles (1982) maps, which approximates to $E(V-I)\sim0.04$
mag via the empirical relation $E(V-I)=1.31 E(B-V)$ (e.g., Mackey \&
Gilmore 2003a,b). MPG06 obtained metallicity and reddening values of
[Fe/H]~$=-0.97\pm0.10$ dex and $E(V-I)=0.04\pm0.02$ mag using the
method of Sarajedini (1994). All of these values are consistent with
our isochrone-fitting result of $E(V-I)=0.03\pm0.01$ mag (using
[Fe/H]~$=-0.97$ dex), which proves that there are no significant
errors in our photometric transformation from the ACS/WFC system to
the Johnson-Cousins $V$ and $I$ system.

We obtained $(m-M)_V=18.30\pm0.06$ mag in this paper, which is larger
than the value from MPG06 ($18.11\pm0.09$ mag) and smaller than the
standard LMC distance modulus ($18.50\pm0.09$ mag, see e.g., Gratton
et al. 2003). In terms of the linear distance, ESO~121--SC03 in our
work is $\sim$ 10 per~cent closer to us than the centre of the LMC,
but $\sim$ 10 per~cent more distant than suggested by the distance
modulus of MPG06. Adopting the optical centre of the LMC at
$\alpha=05^{\rm h}20^{\rm m}56^{\rm s}$ and $\delta=-69^\circ28'41''$
(Bica et al. 1996), ESO~121--SC03 lies at a projected angular
separation of $\sim10^\circ$. Adopting the distance modulus in our
work, the linear distance between the centre of the LMC and
ESO~121--SC03 is $\sim9.5$ kpc. This value is smaller than that in
MPG06 (11.5 kpc), but still agrees with the conclusion in MPG06 that
ESO~121--SC03 is one of the most remote known LMC star clusters.

\section{The Blue Straggler Population}

As shown in Fig.~\ref{fig3}, ESO~121--SC03 is a ``clean'' cluster and
includes an obvious population of BS candidates, which is very
advantageous for our main aim of analysing the BS contributions to the
ISED of the cluster. However, we still need to be very careful in the
selection of BS members, since we do not have access to classical
membership probability assessment methods (i.e., proper motions and/or
radial velocities) that can reliably remove field stars from the
CMD. In this paper, we will try to reduce as much as possible any
contamination to the BS population of the cluster due to field stars
and improper classifications due to possible member stars in the MSTO
region. Two measurements were adopted to avoid exaggerating the BS
population in the cluster.

First, to properly account for the photometric and systematic errors,
we defined a BS region in the CMD using three boundary limits: $V_B,
(V-I)_B$, and the theoretical ZAMS, as shown in Fig.~\ref{fig3},

\begin{eqnarray}
V_{\rm B} &=& V_{\rm TO} - 0.10 \qquad {\rm and}\nonumber \\
(V-I)_{\rm B} &=& (V-I)_{\rm TO} - 0.141 \quad .
\end{eqnarray}

The offset of 0.10 mag in $V_{\rm B}$ relative to the MSTO of the
cluster was adopted as an empirical estimate of the photometric and
systematic errors for {\sl HST}/ACS WFC observations (Sirianni et
al. 2005); the offset of 0.141 mag in $(V-I)_{\rm B}$ was obtained
similarly.

The theoretical ZAMS is represented by the MS ridge line of the
youngest isochrone ($\log(\mbox{age yr}^{-1})=6.60$) in the
Padova1994 library (Bertelli et al. 1994). We have taken this
approach mainly because there is no such an isochrone in the
Padova2000 library: the updated equation of state adopted in
Padova2000 is only sensitive to the evolution of stars with initial
masses from 0.15 to 7 M$_{\odot}$ (Girardi et al. 2000). Therefore,
the age coverage is $\log(\mbox{age yr}^{-1})=[7.80, 10.25]$ in
Padova2000 (a 7 M$_{\odot}$ star is still on the MS at
$\log(\mbox{age yr}^{-1})=6.60$). Meanwhile, the youngest isochrone
fully matches the requirement for a ZAMS in this paper, where it
serves as one of the boundaries for the identification of BSs. The
ZAMS is not involved in fitting the BS positions in the CMD -- for
the fitting we use only the Padova2000 isochrones.

Secondly, the obvious central concentration of BSs has been widely
observed in stellar systems (e.g., Lee et al. 2003 for the Sextans
dwarf spheroidal galaxy; Ferraro et al. 2003 for GCs). Meanwhile,
the radial distributions of BSs are observed to be bimodal in some
clusters (i.e., strongly peaked in the cluster centre, decreasing at
intermediate radii and rising again at greater distances), as shown
by, e.g., Ferraro et al. (1997) for M3, Ferraro et al. (2004) for
47~Tuc, Sabbi et al. (2004) for NGC~6752, and Warren et al. (2006)
for M5. This behaviour has been suggested as probably due to the
relative efficiency of two major BS formation mechanisms: the
mass-transfer scenario should mainly populate the BS population in
the lower-density outskirts of GCs, and the collisional scenario is
the most probable formation mechanism in the high-density inner
regions (e.g., Lanzoni et al. 2007; and references therein).

\begin{figure}
\includegraphics[width=8cm]{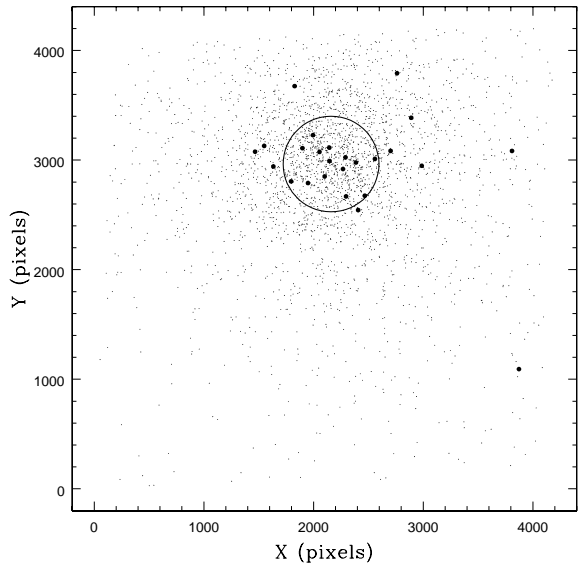}
\caption{Pixel coordinates of all 2824 detected stars in
ESO~121--SC03. The large circle marks the cluster's half-light radius
($R_{\rm HL}=21.8$ arcsec). The solid circles are the
BSs.}\label{fig4}
\end{figure}

Since there are no membership probability measurements available for
our cluster, the BSs identified by their loci in the CMD are all
considered as BS member candidates, and they are divided into two
cases in this paper: (i) BSs inside the half-light radius ($R_{\rm
HL}$) of the cluster. This provides a conservative estimate of the BS
contribution to the total light of the cluster, and these BSs are
marked with open circles in Fig.~\ref{fig3}; (ii) all BSs in the
cluster (pentagons in Fig.~\ref{fig3} and solid circles in
Fig.~\ref{fig4}). This brackets the maximum alteration to the
conventional SSP model for the cluster due to BSs. There are 14 BSs in
case (i) among a total of 25 BSs in case (ii).

Incompleteness corrections were taken into account for the
identification of both the BS population and the cluster's $R_{\rm
HL}$. Comparing the BS selection box in Fig.~\ref{fig3} with the
completeness function in Fig.~\ref{fig1}, we found that the level of
incompleteness was not significant at the luminosity level of the
BSs. The integrated correction for the number of the entire BS
population is only one BS, but one bright and/or blue BS may cause a
significant difference to the modification of the integrated spectral
properties of the cluster. Therefore, instead of taking the risk of
putting an artificial BS somewhere in the BS region, we decided to
keep using the observed BS candidates, and we assumed that the full
sample of BS candidates detected in the cluster corresponded to its
entire BS population. After the incompleteness correction, we derived
the cluster's $R_{\rm HL}$, $R_{\rm HL} =21.8\pm1.0$~arcsec, shown as
a large circle in Fig.~\ref{fig4}.

\section{Modifications to the integrated spectral properties}

\subsection{SSP model reconstruction~--~the synthetic ISED}

The synthetic ISED of the cluster including the BS contribution was
constructed by assuming that the stellar population of the cluster
is made up of two components, i.e., an SSP component and a BS
component. The ISEDs of these two components were derived
separately, and the summation of the ISEDs of these two components
gives the new ISED of the cluster. The resulting synthetic ISED is
then treated as the ISED of the true SSP corresponding to the
parameters of the cluster. The contributions to the ISED of the
cluster due to BSs are measured by comparing the integrated
properties between the conventional ISED (i.e., the ISED of the SSP
component) and the synthetic ISED.

To build the ISED of the BS component, the spectrum of each single
BS must be derived from the observed photometric data. To this end,
we used Padova2000 isochrones of the same metallicity but younger
ages than the cluster's SSP to match the locus in the CMD of each BS
individually. The parameters of effective temperature ($T_{\rm
eff}$) and surface gravity ($\log g$) (roughly equivalent to
luminosity and mass) for each BS can be determined by interpolating
between the two closest isochrones representative of the BS. Based
on the $T_{\rm eff}$ and $\log g$ determined in this way, a
theoretical spectrum from the stellar spectral library of Lejeune et
al. (1997) was assigned to the BS. The theoretical spectra were also
interpolated, so that the same metallicity as that of the cluster
was adopted. This process was performed individually for all BSs in
the cluster, and the ISED of the BS component is given by adding up
the individual spectra of the BSs directly.

\begin{equation}
F_{BS}=\sum_{i=1}^{N_{BS}}f^i_{BS} \quad ,
\end{equation}

A significant amount of theoretical work has focused on BS formation
models. BSs from equal-mass MS star collisions (Benz \& Hills 1987)
may suffer from huge mass loss and undergo complete chemical
composition mixing -- the unclear clock will, in effect, be reset to
a ZAMS star. Unequal-mass MS star collisions (Benz \& Hills 1992)
may cause much less mass loss, but also violent mixing in the
atmosphere of the more massive star -- this may revert the
observational parameters of the more massive star back to those of a
helium-rich MS star. Meanwhile, Lombardi et al. (1995, 2002) argued
that BSs from stellar collisions would not be expected to have
either a significant helium enhancement in the outer envelope, nor
replenished fresh hydrogen fuel in their cores. Detailed spectral
analysis of BSs and turn-off region MS stars in M67 (e.g., Mathys
1991; Shetrone \& Sandquist 2000) led to the conclusion that the Fe
and Ca abundances of BSs are very close to those of the MS stars in
the cluster. Liu et al. (in prep.) conclude that most BSs exhibit
similar properties to those of regular MS stars, based on a detailed
analysis of low-resolution spectra of all BSs in M67.

It is clear that rotational support and the stellar chemical profile
are very important, because they have profound effects on stellar
evolution and on the remnant's position in a CMD (Sills et al.
2000), to which our method of obtaining the BS spectrum is certainly
sensitive. However, until either a well-established theory emerges
that can predict the actual internal structure of all BSs in a CMD,
or high-resolution spectra are available for all BSs in a CMD, the
theoretical approaches followed in this series of papers (XD05,
XDH07, and the present paper) are fully adequate, as long as we
restrict ourselves to low-resolution spectroscopic analysis.

The main goal of this series of papers is to remedy the conventional
SSP models empirically by analyzing the contributions of the
observed BSs quantitatively. To reach this goal, what is really
needed is a quantification of the relative modifications to
conventional SSP models when the same basic ingredients (i.e.,
isochrone library, spectral library, and IMF) are used to describe
both the conventional SSP and the BS contribution. As discussed in
the previous sections, the positions of BSs in the observed CMD are
fitted with Padova2000 isochrones, and the BS spectra are extracted
from the Lejeune et al. (1997) stellar spectral library. Therefore,
the conventional SSP models (BC03) based on the same libraries and
using a Salpeter (1955) IMF are adopted to represent the properties
of the SSP components in star clusters. We interpolated the ISEDs
across a grid of age and metallicity to construct the ISED of the
SSP component so that they match the parameters of the cluster.

For an SSP of age $t$ and metallicity $Z$, assuming an IMF
$\phi(m)$, the ISED of the SSP component is
\begin{equation}
F_{\rm SSP}(\lambda,t,Z)=A\int_{m_l}^{m_u}\phi(m)f(\lambda,m,t,Z){\rm
d}m \quad ,
\end{equation}
where $f(\lambda,m,t,Z)$ is the flux of a single star of mass $m$,
age $t$, and metallicity $Z$; $m_u$ and $m_l$ are the upper and
lower integration limits in mass, respectively. $A$ is a
normalization constant, which is used to restore the flux of the SSP
component to its real intensity.

The ISEDs of BC03 SSP models are all normalized to a total mass of 1
M$_\odot$ in stars, at $t = 0$. Therefore, we need to calibrate the
ISEDs of the BS and SSP components similarly. In order to do so, we
count the number of stars in the CMD in the magnitude interval
between the MSTO and the point at two magnitudes below the MSTO on
the MS. This number is defined as $N_2$ (Ahumada \& Lapasset 1995).
$N_2$ is used to calculate the normalization constant $A$ in Eq.
(3). By using a Salpeter IMF, $A$ can be derived as follows:

\begin{equation}
N=A\int_{m_1}^{m_2}m^{-2.35} {\rm d} m=N_2 \quad ,
\end{equation}
where $m_2$ and $m_1$ are, respectively, the stellar masses at the
MSTO and at the point 2 magnitudes below the MSTO. To ensure the
integrity of the stellar number counts in this range, the photometry
should be sufficiently deep. The photometric observational data used
in this paper meets this requirement. Corrections for incompleteness
in this mass range are also applied. For a detailed description of
the model construction procedure, see XD05 and XDH07.

Incidentally, the choice of the IMF also introduces uncertainties in
the SSP's integrated properties, but it is unimportant for the
discussion in this series of papers. The conventional SSP models
using a Salpeter IMF are only used as a reference to the ISEDs of
the SSP components, what is really important in the current context
is the contributions of BSs to the ISED; the properties of the BSs
in our cluster are determined individually, and independently of the
adopted IMF properties.

\subsection{Modification to the ISED}

\begin{figure}
\includegraphics[width=8cm]{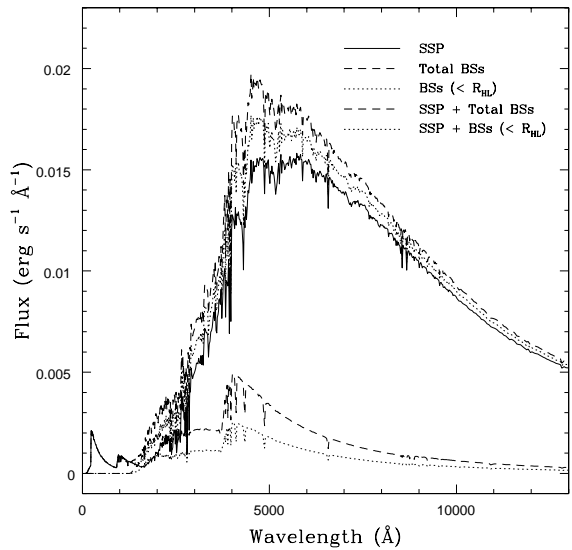}
\caption{ISED modifications. The BS contributions are presented for
two cases. The solid line is the ISED of the SSP component. The thin
dotted line is the ISED of the BS component for case (i): BSs within
$R_{\rm HL}$. The thin dashed line is the ISED of the BS component
for case (ii): all BSs in the cluster. The heavy dotted line is the
synthetic ISED of the SSP component and the case-(i) BS component.
The heavy dashed line is the synthetic ISED of the SSP component and
the case-(ii) BS component.}\label{fig5}
\end{figure}

According to their loci in the CMD (Fig.~\ref{fig3}), BSs are the
bluest cluster members, with extraordinary luminosities, and thus they
may cause very significant modifications to the conventional view of
the integrated spectral properties of a cluster. The contributions of
BSs to a cluster's ISED are demonstrated in Fig.~\ref{fig5} for two
cases: (i) considering only the BSs inside $R_{\rm HL}$; and (ii) all
BSs in the cluster.

In Fig.~\ref{fig5}, the solid line is the ISED of the conventional SSP
model with an isochrone-fitted (regarded as ``true'') age and
metallicity representative of that of the cluster, i.e., the ISED of
the SSP component. The thin dotted line is the ISED of the BS
component for case (i); the thin dashed line is the ISED of the BS
component for case (ii). The heavy dotted line is the synthetic ISED
of the SSP component and the case-(i) BS component; the heavy dashed
line is the synthetic ISED of the SSP component and the case-(ii) BS
component. Although they are not dominated by BSs, both synthetic
ISEDs show significant enhancements across a wide range of UV, blue
and visual wavelengths. The synthetic ISEDs become undoubtedly hotter
than the ISED of the conventional SSP (solid line in
Fig.~\ref{fig5}). Taking the values of the flux at 5510 {\AA} as an
example, the energy enhancements will be, respectively, $\sim10$ and
$\sim 18$ per cent if we include the case-(i) and case-(ii) BS
components.

\subsection{Modification to the broad-band colours}

This series of papers (this paper, XD05, and XDH07) has indicated
that BSs in general stellar populations should be recognized and
treated with special care when dealing with EPS studies of galaxies.

In addition to direct ISED analysis and spectral-index applications,
photometric observations using broad-band filters are more frequently
used than spectrophotometry when trying to understand stellar
populations in distant and therefore often very faint galaxies. As
shown in Fig.~\ref{fig5}, the BS component enhances the ISED of the
conventional SSP across a wide range of wavelengths, especially in the
UV and blue. This tends to make the stellar population appear younger
and/or more metal poor according to conventional SSP models, since
both modifications can make the colours bluer.

Four broad-band colours, $(U-B), (B-V), (V-R)$, and $(V-I)$, are
taken as probes in this paper to quantitatively measure the BS
contributions for ESO~121--SC03. The colours are obtained by
convolving the ISEDs with the corresponding filter response curves.
The results are listed in Table~\ref{tab1}. The colours on the first
line are from the ISED of the cluster's SSP component, those on the
second line are from the synthetic ISED of the SSP and the case-(i)
BS component, and those on the third line are from the synthetic
ISED of the SSP and the case-(ii) BS component. Based on the results
in Table~\ref{tab1}, we conclude that the broad-band colours are all
modified, to some extent, with respect to the conventional models
(line 1) owing to the presence of BSs. The influence of the BSs
becomes more pronounced toward bluer wavelengths. For $(U-B)$, the
colour modification can be up to $\sim 50$ per cent. For $(B-V)$ and
$(V-R)$, the modifications are $\sim10$ per cent. Even for $(V-I)$,
our reddest colour, the changes are still more than 5 per cent.
Although these modifications are specific to the case of ESO
121-SC03, i.e., for [Fe/H]~$=-0.97$ dex and an age of 8.9 Gyr, this
example serves as a warning for the more general use of broad-band
colours to derive cluster ages and metallicities based on
theoretical SSP modeling.

\begin{table}
\centering
\begin{minipage}{140mm}
\caption{The broad-band colours of the cluster for different
cases.}\label{tab1}
\begin{tabular}{lcccc}
\hline
Component & $(U-B)$ & $(B-V)$ & $(V-R)$ & $(V-I)$ \\
          &  (mag)  &  (mag)  &  (mag)  &  (mag)  \\
\hline
SSP                         & 0.101 & 0.744 & 0.516 & 0.993 \\
SSP + BSs $\leq R_{\rm HL}$ & 0.091 & 0.683 & 0.487 & 0.945 \\
SSP + all BSs               & 0.045 & 0.635 & 0.453 & 0.905 \\
\hline
\end{tabular}
\end{minipage}
\end{table}

Based on the modifications to the ISED and the broad-band colours
calculated as examples in this paper, which are much greater than the
observational errors, we strongly caution that conventional SSP models
should be used carefully in evolutionary population synthesis of
galaxies, at least for the intermediate and old population-I stars. It
is difficult to trace the BS population in very young star clusters,
and the old population-II star clusters usually have very extended HBs
which, instead of the BSs, will dominate the integrated light at blue
wavelengths.

\section{Underestimates of ages and/or metallicities by the
conventional SSP models}

In the previous section we showed that the ISED and broad-band colours
of conventional SSPs (for the case of ESO~121--SC03) are modified
dramatically by BS components, and the modifications will introduce
sizeable uncertainties if conventional SSP models are used to obtain
the basic physical parameters of a stellar population. Again, taking
ESO~121--SC03 as an example, we try to quantify the uncertainties by
fitting the synthetic ISED of the cluster using conventional SSP
models. We focus on the age and metallicity, the two most important
parameters of a stellar population.

As presented in Fig.~\ref{fig5}, one of the main effects of BSs is
to make the synthetic ISED hotter with respect to that of the
conventional models, i.e., the spectra get enhanced more
significantly towards shorter wavelengths. Technically, this hotter
ISED can still be well fitted (or, say, misunderstood) by
conventional SSP models of either younger age or lower metallicity,
since both options make the ISED hotter. Therefore, we regard the
synthetic ISED of ESO~121--SC03 as the real ISED (i.e., the ISED
that will be observed spectroscopically), and we fit it with an ISED
of a conventional SSP model. The standard deviation ($\sigma$) is
used to identify the best-fitting result. The differences between
the real cluster parameters and the best-fitting results are used to
discuss the uncertainties that intrinsically exist in the
application of the EPS method.

We emphasize here that instead of building the actual ISED of a
specific cluster, what we are really interested in is to discuss the
potential uncertainties introduced when the BS component is
completely ignored. Thus, although the synthetic ISED is not exactly
the real ISED of the cluster, the method and corresponding results
are effective approaches to the problem.

\subsection{Uncertainties in age}

Fig.~\ref{fig6} shows the uncertainties in age: we fit the synthetic
cluster ISED with the conventional SSP models, keeping the
metallicity fixed while lowering the age values. The left-hand panel
shows the results for the synthetic ISED of the SSP component plus
the case-(ii) BS component (all BSs); the right-hand panel is the
same but for the SSP component plus the case-(i) BS component (BSs
within $R_{\rm HL}$). In the top panels, the solid line is the
synthetic cluster ISED, the dotted line is the best-fitting ISED
based on conventional SSP models, and the dashed line is the ISED of
the SSP component. Plotting the SSP component's ISED in the figure
is done to show the differences between the real ISED and the
conventional model ISED with the true parameters, and to show that
the SSP component's ISED does not correctly represent the
observations. The conventional ISEDs are normalized to the real ISED
at a wavelength of 5500 {\AA}. In the bottom panels, the residuals
between the synthetic ISED and the ISEDs of two conventional SSPs
and the $3\sigma$ regions are shown, respectively, using matching
line styles.

Based on the results shown in Fig.~\ref{fig6} and assuming that the
observed ISED is the only data available for stellar population
analysis, the fitting with the conventional SSP models will result
in a best-fitting age of 3.4 Gyr, which is more than 60 per cent
younger than the real age of the cluster, 8.9 Gyr, determined from
isochrone fits to the CMD; even considering only the BSs within
$R_{\rm HL}$, the best-fitting age of 5.4 Gyr still suffers from an
non-negligible difference of $\sim40$ per cent. In the bottom
panels, the $\sigma$ values are indeed small for all cases --
approximately at the level of only one per cent of the cluster's
flux, but the $\sigma$ value according to the best-fitting
conventional ISED is always $\sim \frac{2}{3} - \frac{3}{4}$ smaller
than that from the SSP component of the cluster alone, which implies
that the EPS method will probably lead to an incorrect best-fitting
age in unresolved conditions.

\subsection{Uncertainties in metallicity}

\begin{figure}
\includegraphics[width=8cm]{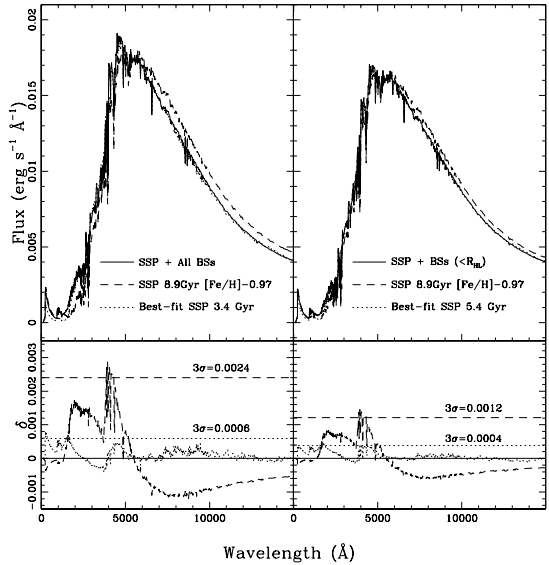}
\caption{Fits of the synthetic ISED of ESO~121--SC03 with conventional
SSP models~--~uncertainties in the age determination. The left-hand
panel is for the synthetic ISED of the SSP component and all BSs. The
right-hand panel is for the synthetic ISED of the SSP component and
the BSs inside $R_{\rm HL}$. In the top panels, the solid line is the
synthetic cluster ISED (real ISED), the dotted line is the
best-fitting ISED based on the conventional SSP models, and the dashed
line is the ISED of the SSP component. The model ISEDs are normalised
to the real ISED at a wavelength of 5500 {\AA}. In the bottom panels,
the differences between the synthetic ISED and the ISEDs of two
conventional SSPs and the $3\sigma$ regions are shown, respectively,
using matching line styles.}\label{fig6}
\end{figure}

\begin{figure}
\includegraphics[width=8cm]{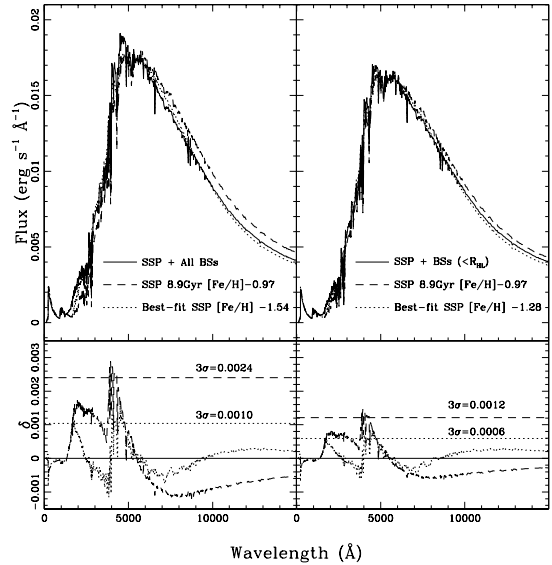}
\caption{Fits of the synthetic ISED of ESO~121--SC03 with conventional
SSP models~--~uncertainties in the metallicity determination. Figure
coding is as for Fig. \ref{fig6}.}\label{fig7}
\end{figure}

In a similar way as in the previous section, Fig.~\ref{fig7} shows the
uncertainties in metallicity in evolutionary population synthesis
using the conventional SSP models. The meaning of the symbols and line
styles in this figure is similar to that in Fig.~\ref{fig6}. The
results are also similar: the metallicities (expressed in [Fe/H]) are
significantly underestimated (i.e., more metal poor) as derived from
the conventional SSP models. The underestimates will be $\sim60$ and
$\sim 30$ per cent, respectively, if we include the cases (ii) and (i)
BS components. The detailed results are listed in
Table~\ref{tab2}. Taking the modifications in age as an example, the
fit uncertainty is calculated as
\begin{equation}
\Delta_{\rm fit} = \frac{(\rm age)_{\rm real}-(\rm age)_{\rm
fit}}{(\rm age)_{\rm real}}.
\end{equation}

As shown in Figs.~\ref{fig6} and \ref{fig7}, the conventional SSP
models can fit perfectly most of the features of the observed ISED,
if we leave the parameters free. However, such good fits do not mean
that we have obtained good determinations of the real physical
parameters in unresolved conditions. Applications of conventional
SSP models in EPS studies may therefore seriously suffer from the
uncertainties addressed in this paper. This is true at least for the
stellar populations corresponding to the star clusters that we have
thus far analyzed.

\begin{table*}
\centering
\begin{minipage}{140mm}
\caption{Fit Uncertainties for ESO~121--SC03}\label{tab2}
\begin{tabular}{cccccc}
\hline
 {} & \multicolumn{2}{c}{SSP + All BSs} &  &
\multicolumn{2}{c}{SSP + BSs ($\leq R_{\rm HL}$}) \\
\cline{2-3} \cline{5-6} \\
{Real Parameter} & Age = 8.9 Gyr & [Fe/H] $=-0.97$ & {} &
Age = 8.9 Gyr & [Fe/H] $=-0.97$\\
\hline
Model Fit              & 3.4  & $-$1.54 & & 5.4  & $-$1.28 \\
Uncertainty (per cent) & 62   & 59      & & 40   & 32 \\
\hline
\end{tabular}
\end{minipage}
\end{table*}

\section{Summary and conclusions}
In this paper, the LMC star cluster ESO~121--SC03 is taken as a
specific cluster to show in detail the technique and procedure we
have developed to assess the potential importance of BSs to the
integrated properties of unresolved star clusters: data reduction,
isochrone fitting, identification of the BS population, synthetic
ISED construction, and the concept of detecting BS contributions to
conventional SSP models using integrated light properties.

We construct the synthetic ISED of the cluster by decomposing the
stellar population of the cluster into two components: the SSP
component composed of all the member stars that can be well fitted
by an isochrone of single stars; and the BS component defined by the
BSs in the cluster. The synthetic ISED of the sum of these two
components is regarded as the real ISED of the cluster. The
contributions to the ISED of the cluster owing to the BS component
are considered for two cases. For case (i), only BSs inside $R_{\rm
HL}$ are used, which provides a conservative estimate of the
modification caused by the presence of BSs; while for case (ii), all
BSs in the cluster are included, which renders the maximum likely
alteration to the conventional results. These two cases bracket a
range of changes to the conventional SSP models caused by the BS
population in a real cluster.

Integrated spectral properties are used to measure the modifications
to the conventional SSP due to BSs. By the inclusion of BSs, the
synthetic ISED of ESO~121--SC03 is greatly enhanced toward shorter
wavelengths, and all UV and optical broad-band colours become bluer.
Both of these observations can make the population appear a lot
younger or more metal poor than would be derived from fitting
conventional SSP models. The modifications are quantified by fitting
the synthetic ISED with conventional SSP models. For the synthetic
cluster ISED including only BSs inside $R_{\rm HL}$, the age
underestimate is at the level of $\sim40$ per cent, and the
metallicity underestimate is $\sim30$ per cent. For the synthetic
ISED that includes all BSs identified in the cluster, the
underestimates of both age and metallicity increase to $\sim60$ per
cent.

In view of the common occurrence of BSs in MC star clusters (e.g.,
Johnson et al. 1999; Alcaino et al. 2003), and given that our
results for the comparatively rich MC clusters (at least, with
respect to the Galactic OCs) are statistically robust, we have
reason to believe that the BS effect discussed in this paper for
ESO~121--SC03 may not be an exception. BS contributions (to the
conventional SSPs) measured based on other MC star clusters are
expected to be at a similar level as for ESO~121--SC03, and likely
follow the same trend as for our previous results based on Galactic
OCs covering a wide age range (Deng et al. 1999; XD05; XDH07): the
clusters' age and/or metallicity estimates will be underestimated
significantly, by a factor of 2--4, based on fits of SSPs to either
spectra or broad-band colours. In a subsequent paper, we will
discuss the equivalent results for the entire MC star cluster
sample, i.e., based on the application of the method discussed here,
which will significantly extend the parameter coverage and the
robustness of this conclusion.

In order to constrain the ISEDs of real stellar populations in a
reliable way and to provide the community with a set of modified SSP
models, extending our work on cluster samples to the MC rich star
clusters is now in progress. In the context of our ultimate goal, we
are currently still at the stage of information collection. To
correct the conventional SSP models empirically, a sufficient
working sample of clusters covering large ranges of age and
metallicity, and of different dynamical environments, is definitely
needed. Such a sample will provide statistical results of the BS
contributions, and reliable information of BS populations in SSPs,
such as the specific frequency of BSs (BS numbers in an SSP), and
luminosity and colour functions of BSs (BS positions in a CMD).
These will, we expect, eventually lead to the final solution to this
problem, when we can construct the BS-corrected SSP models which can
then be used in EPS studies in unresolved conditions.

\section*{Acknowledgments}
YX, LD and ZH would like to thank the Chinese National Science
Foundation for support through grants 10573022, 10778719, 10333060,
10521001, 10433030, and the Ministry of Science and Technology of
China through grant 2007CB815406. YX acknowledges financial support
from the Royal Society in the form of a ``Sino-British Fellowship
Trust Award''. YX is also grateful to the International Astronomical
Union for travel support from Beijing to the UK. RdG acknowledges
partial funding from the Royal Society through an International
Joint Project grant, as well as support for a high-level UK-China
Science Network, funded by the Office of Science and Innovation of
the UK government. ADM is supported by a Marie Curie Excellence
Grant from the European Commission under contract
MCEXT-CT-2005-025869.

This paper is based on observations made with the NASA/ESA
\textit{Hubble Space Telescope}, obtained at the Space Telescope
Science Institute, which is operated by the Association of
Universities for Research in Astronomy, Inc., under NASA contract NAS
5-26555. These observations are associated with programme \#9891.

\bsp

\label{lastpage}


\begin{thebibliography}{99}
\bibitem[\protect\citeauthoryear{Ahumada \& Lapasset}{1995}]{b}
Ahumada J.A., Lapasset E., 1995, A\&AS, 109, 375
\bibitem[\protect\citeauthoryear{Ahumada \& Lapasset}{2007}]{b}
Ahumada J.A., Lapasset E., 2007, A\&A, 463, 789
\bibitem[\protect\citeauthoryear{Alcaino et al.}{2003}]{b} Alcaino G.,
Alvarado F., Borissova J., Kurtev R., 2003, A\&A, 400, 917
\bibitem[\protect\citeauthoryear{Bekki et al.}{2004}]{b}
Bekki K., Beasley M.A., Forbes D.A., Couch W.J., 2004, ApJ, 602, 730
\bibitem[\protect\citeauthoryear{Benz \& Hills}{1987}]{b} Benz W.,
Hills J.G., 1987, ApJ, 323, 614
\bibitem[\protect\citeauthoryear{Benz \& Hills}{1992}]{b} Benz W.,
Hills J.G., 1992, ApJ, 389, 546
\bibitem[\protect\citeauthoryear{Bertelli et al.}{1994}]{b} Bertelli
G., Bressan A., Chiosi C., Fagotto F., Nasi E., 1994, A\&AS, 106,
275
\bibitem[\protect\citeauthoryear{Bica et al.}{1996}]{b} Bica E.,
Clari\'{a} J.J., Dottori H., Santos Jr.J.F.C., Piatti A.E., 1996,
ApJS, 102, 57
\bibitem[\protect\citeauthoryear{Bica et al.}{1998}]{b} Bica E.,
Geisler D., Dottori H., Clari\'{a} J.J., Piatti A.E., Santos
Jr.J.F.C., 1998, AJ, 116, 723
\bibitem[\protect\citeauthoryear{Bruzual \& Charlot}{2003}]{b}
Bruzual A. G., Charlot S., 2003, MNRAS, 344, 1000 (BC03)
\bibitem[\protect\citeauthoryear{Burstein \& Heiles}{1982}]{b}
Burstein D., Heiles C., 1982, AJ, 87, 1165
\bibitem[\protect\citeauthoryear{Carney et al.}{2005}]{b} Carney B.W.,
Latham D.W., Laird J.B., 2005, AJ, 129, 466
\bibitem[\protect\citeauthoryear{Da~Costa}{1991}]{b}
Da~Costa G.S., 1991, in IAU Symp. 148, The Magellanic Clouds, Eds.
Haynes R., Milne D., (Dordrecht: Kluwer), p183
\bibitem[\protect\citeauthoryear{de Grijs \& Anders}{2006}]{b} de~Grijs
R., Anders P., 2006, MNRAS, 366, 295
\bibitem[\protect\citeauthoryear{Deng et al.}{1999}]{b} Deng L.,
Chen R., Liu X.S., Chen J.S., 1999, ApJ, 524, 824
\bibitem[\protect\citeauthoryear{Dirsch et al.}{2000}]{b} Dirsch B.,
Richtler T., Gieren W.P., Hilker M., 2000, A\&A, 360, 133
\bibitem[\protect\citeauthoryear{Dolphin}{2000}]{b} Dolphin A.E.,
2000, PASP, 112, 1383
\bibitem[\protect\citeauthoryear{Ferraro et al.}{2004}]{b} Ferraro
F.R., Beccari G., Rood R.T., Bellazzini M., Sills A., Sabbi E.,
2004, ApJ, 603, 127
\bibitem[\protect\citeauthoryear{Ferraro et al.}{1997}]{b} Ferraro
F.R., Paltrinieri B., Fusi~Pecci F., Cacciari C., Dorman B., Rood
R.T., Buonanno R., Corsi C.E., Burgarella D., Laget M., 1997, A\&A,
324, 915
\bibitem[\protect\citeauthoryear{Ferraro et al.}{2003}]{b} Ferraro
F.R., Sills A., Rood R.T., Paltrinieri B., Buonanno R., 2003, ApJ,
588, 464
\bibitem[\protect\citeauthoryear{Geisler et al.}{1997}]{b}
Geisler D., Bica E., Dottori H., Clari J.J., Piatti A.E., Santos
Jr.J.F.C., 1997, AJ, 114, 1920
\bibitem[\protect\citeauthoryear{Girardi et al.}{2000}]{b} Girardi
L., Bressan A., Bertelli G., Chiosi C., 2000, A\&AS, 141, 371
\bibitem[\protect\citeauthoryear{Gratton et al.}{2003}]{b} Gratton
R.G., Bragaglia A., Carretta E., Clementini G., Desidera S.,
Grundahl F., Lucatello S., 2003, A\&A, 408, 529
\bibitem[\protect\citeauthoryear{Johnson et al.}{1999}]{b} Johnson J.A.,
Bolte M., Stetson P.B., Hesser J.E., Somerville R.S., 1999, ApJ,
527, 199
\bibitem[\protect\citeauthoryear{Lanzoni et al.}{2007}]{b} Lanzoni
B., Dalessandro E., Ferraro F.R., Mancini C., Beccari G., Rood R.T.,
Mapelli M., Sigurdsson S., 2007, ApJ, 663, 267
\bibitem[\protect\citeauthoryear{Lee et al.}{2003}]{b} Lee M.G., et
al., 2003, AJ, 126, 2840
\bibitem[\protect\citeauthoryear{Lejenue et al.}{1997}]{b} Lejeune
Th., Cuisinier F., Buser R., 1997, A\&AS, 125, 229
\bibitem[\protect\citeauthoryear{Liu et al.}{2008}]{b} Liu G., Deng
L., Chavez M.D., et al., 2008, in preparation
\bibitem[\protect\citeauthoryear{Lombardi et al.}{1995}]{b} Lombardi
J.C.Jr., Rasio F.A., Shapiro S.L., 1995, ApJ, 445L, 117
\bibitem[\protect\citeauthoryear{Lombardi et al.}{2002}]{b} Lombardi
J.C.Jr., Warren J.S., Rasio F.A., Sills A., Warren A.R., 2002, ApJ,
568, 939
\bibitem[\protect\citeauthoryear{Mackey \& Broby~Nielsen}{2007}]{b}
Mackey A.D., Broby~Nielsen P., 2007, MNRAS, 379, 151
\bibitem[\protect\citeauthoryear{Mackey \& Gilmore}{2004}]{b} Mackey
A.D., Gilmore G.F., 2004, MNRAS, 352, 153
\bibitem[\protect\citeauthoryear{Mackey, Payne \& Gilmore}{2006}]{b} Mackey
A.D., Payne M.J., Gilmore G.F., 2006, MNRAS, 369, 921 (MPG06)
\bibitem[\protect\citeauthoryear{Mackey \& Gilmore}{2003a}]{b}
Mackey A.D., Gilmore G.F., 2003a, MNRAS, 338, 85
\bibitem[\protect\citeauthoryear{Mackey \& Gilmore}{2003b}]{b}
Mackey A.D., Gilmore G.F., 2003b, MNRAS, 338, 120
\bibitem[\protect\citeauthoryear{Mapelli et al.}{2007}]{b} Mapelli M.,
Ripamonti E., Tolstoy E., Sigurdsson S., Irwin M.J., Battaglia G.,
2007, MNRAS, 380, 1127
\bibitem[\protect\citeauthoryear{Mateo et al.}{1986}]{b} Mateo M.,
Hodge P., Schommer R.A., 1986, ApJ, 311, 113
\bibitem[\protect\citeauthoryear{Mathys}{1991}]{b} Mathys G., 1991,
A\&A, 245, 467
\bibitem[\protect\citeauthoryear{Parmentier \& de~Grijs}{2007}]{b}
Parmentier G., de~Grijs R., 2007, MNRAS, accepted (arXiv:0710.3477)
\bibitem[\protect\citeauthoryear{Piotto et al.}{2004}]{b} Piotto G., De
Angeli F., King I.R., Djorgovski S.G., Bono G., Cassisi S., Meylan
G., Recio-Blanco A., Rich R.M., Davies M.B., 2004, ApJ, 604, L109
\bibitem[\protect\citeauthoryear{Portegies~Zwart et al.}{1997b}]{b}
Protegies~Zwart S.F., Hut P., McMillan S.L.W., Verbunt F., 1997b,
A\&A, 328, 143
\bibitem[\protect\citeauthoryear{Portegies~Zwart et al.}{1997a}]{b}
Protegies~Zwart S.F., Hut P., Verbunt F., 1997a, A\&A, 328, 130
\bibitem[\protect\citeauthoryear{Sabbi et al.}{2004}]{b} Sabbi E.,
Ferraro F.R., Sills A., Rood R.T., 2004, ApJ, 617, 1296
\bibitem[\protect\citeauthoryear{Salpeter}{1955}]{b} Salpeter E.E.,
1955, ApJ, 121, 161
\bibitem[\protect\citeauthoryear{Sarajedini}{1994}]{b} Sarajedini
A., 1994, AJ, 107, 618
\bibitem[\protect\citeauthoryear{Shetrone \& Sandquist}{2000}]{b}
Shetrone M.D., Sandquist E.L., 2000, AJ, 120, 1913
\bibitem[\protect\citeauthoryear{Sills et al.}{2005}]{b} Sills A., Adams
T., Davies M.B., 2005, MNRAS, 358, 716
\bibitem[\protect\citeauthoryear{Sills et al.}{2000}]{b} Sills A.,
Pinsonneault M.H., Terndrup D.M., 2000, ApJ, 534, 335
\bibitem[\protect\citeauthoryear{Sirianni et al.}{2005}]{b} Sirianni
M., et al., 2005, PASP, 117, 1049
\bibitem[\protect\citeauthoryear{Tian et al.}{2006}]{b} Tian B., Deng L.,
Han Z.W., Zhang X.B., 2006, A\&A, 455, 247
\bibitem[\protect\citeauthoryear{Warren et al.}{2006}]{b} Warren
S.R., Sandquist E.L., Bolte M., 2006, ApJ, 648, 1026
\bibitem[\protect\citeauthoryear{Xin \& Deng}{2005}]{b} Xin Y., Deng
L., 2005, ApJ, 619, 824 (XD05)
\bibitem[\protect\citeauthoryear{Xin, Deng \& Han}{2007}]{b} Xin Y., Deng L., Han
Z.W., 2007, ApJ, 660, 319 (XDH07)

\end{thebibliography}
\end{document}